\documentclass[twocolumn]{aastex62}
\usepackage{graphicx}	% Including figure files
\usepackage{amsmath}	% Advanced maths commands
\usepackage{amssymb}
\usepackage{natbib}
\usepackage[frozencache=true]{minted}
\DeclareMathOperator{\sech}{sech}
\usepackage{url}

\graphicspath{{./}{figures/}}

\received{March 28, 2018}
\revised{July 6, 2018}
\revised{August 13, 2018}
\accepted{August 24, 2018}
\shorttitle{Fast orbital parameter estimation}
\shortauthors{Mackereth \& Bovy}
\begin{document}

\title{FAST ESTIMATION OF ORBITAL PARAMETERS IN MILKY-WAY-LIKE POTENTIALS}

\correspondingauthor{J. Ted Mackereth}
\email{J.E.Mackereth@2011.ljmu.ac.uk}

\author[0000-0002-0786-7307]{J. Ted Mackereth}
\affil{Astrophysics Research Institute, Liverpool John Moores University, 146 Brownlow Hill, Liverpool, L3 5RF, United Kingdom}

\author[0000-0001-6855-442X]{Jo Bovy}
\altaffiliation{Alfred P. Sloan Fellow}
\affiliation{Department of Astronomy \& Astrophysics, University of Toronto, 50 St. George Street, Toronto, ON  M5S 3H4, Canada}

%% Note that the \and command from previous versions of AASTeX is now
%% depreciated in this version as it is no longer necessary. AASTeX 
%% automatically takes care of all commas and "and"s between authors names.

%% AASTeX 6.2 has the new \collaboration and \nocollaboration commands to
%% provide the collaboration status of a group of authors. These commands 
%% can be used either before or after the list of corresponding authors. The
%% argument for \collaboration is the collaboration identifier. Authors are
%% encouraged to surround collaboration identifiers with ()s. The 
%% \nocollaboration command takes no argument and exists to indicate that
%% the nearby authors are not part of surrounding collaborations.

%% Mark off the abstract in the ``abstract'' environment. 
\begin{abstract}

Orbital parameters, such as eccentricity and maximum vertical excursion, of stars in the Milky Way are an important tool for understanding its dynamics and evolution, but calculation of such parameters usually relies on computationally-expensive numerical orbit integration. We present and test a fast method for estimating these parameters using an application of the St\"ackel fudge, used previously for the estimation of action-angle variables. We show that the method is highly accurate, to a level of $<1\%$ in eccentricity, over a large range of relevant orbits and in different Milky Way-like potentials, and demonstrate its validity by estimating the eccentricity distribution of the RAVE-TGAS data set and comparing it to that from orbit integration. Using the method, the orbital characteristics of the $\sim 7$ million \emph{Gaia} DR2 stars with radial velocity measurements are computed with Monte Carlo sampled errors in $\sim 116$ hours of parallelised cpu time, at a speed that we estimate to be $\sim 3$ to $4$ orders of magnitude faster than using numerical orbit integration. We demonstrate using this catalogue that \emph{Gaia} DR2 samples a large range of orbits in the solar vicinity, down to those with $r_\mathrm{peri} \lesssim 2.5$ kpc, and out to $r_\mathrm{ap} \gtrsim 13$ kpc. We also show that many of the features present in orbital parameter space have a low mean $z_\mathrm{max}$, suggesting that they likely result from disk dynamical effects.

\end{abstract}

%% Keywords should appear after the \end{abstract} command. 
%% See the online documentation for the full list of available subject
%% keywords and the rules for their use.

%% From the front matter, we move on to the body of the paper.
%% Sections are demarcated by \section and \subsection, respectively.
%% Observe the use of the LaTeX \label
%% command after the \subsection to give a symbolic KEY to the
%% subsection for cross-referencing in a \ref command.
%% You can use LaTeX's \ref and \label commands to keep track of
%% cross-references to sections, equations, tables, and figures.
%% That way, if you change the order of any elements, LaTeX will
%% automatically renumber them.
%%
%% We recommend that authors also use the natbib \citep
%% and \citet commands to identify citations.  The citations are
%% tied to the reference list via symbolic KEYs. The KEY corresponds
%% to the KEY in the \bibitem in the reference list below. 

\section{Introduction}

The orbit integration of stars from their observed 6D phase-space coordinates is an important method for better understanding the kinematic and dynamical properties of galaxies and their stellar populations, but can be time-consuming computationally, especially when large sample sizes are involved. While various methods have been devised for estimating the angle-action variables for given sets of phase-space coordinates \citep[see, e.g.][and references therein]{2016MNRAS.457.2107S}, more conventional orbital parameters such as eccentricity for increasingly large samples in the era of \emph{Gaia} may still offer important insights into the nature of the Milky Way. 

	As an example, orbital eccentricities have already been effective in trying to understand the origins of the thicker disc component in the Galaxy. It was suggested by \citet{2009MNRAS.400L..61S} that the eccentricity distribution as a function of height above the mid-plane could be constraining to thick disc formation models. Calculations of the orbital eccentricity from orbit integration of 31,535 stars from SDSS DR7 were later employed to test this idea, and place constraints on the origin of the thicker disc components \citep{2010ApJ...725L.186D}. The apocenter and pericenter radii of orbits (as well as their angular momenta) have also been employed as a means of finding substructure in the Milky Way disc \citep[the APL space, ][]{2006MNRAS.365.1309H}.
    
    While actions are more fundamental in labelling orbits and perhaps more useful for dynamical modelling (being distinguished from other labels by their adiabatic invariance), we argue that the regular orbit parameters: maximum vertical excursion $z_{\mathrm{max}}$, pericenter and apocenter radius $r_{\mathrm{peri}},r_{\mathrm{ap}}$, and their transformation into orbital eccentricity $e$, can still be useful for kinematics studies and are more naturally related to the orbital configuration of the Galaxy and its relation to its formation history. Used in tandem with computed orbital actions, these orbit labels can aid in the understanding and disentangling of, for example, substructures that are discovered in action space, as the parameters are expressed as physical distances. In this paper, we present a method for the rapid estimation of these orbital parameters analytically, without recourse to numerical orbit integration. This method is a simplified application of the St\"ackel fudge presented by \citet{2012MNRAS.426.1324B} to compute actions and angles for axisymmetric potentials.

\section{Method}
\label{sec:method}

\begin{figure*}
\begin{centering}
\includegraphics[width=0.8\textwidth]{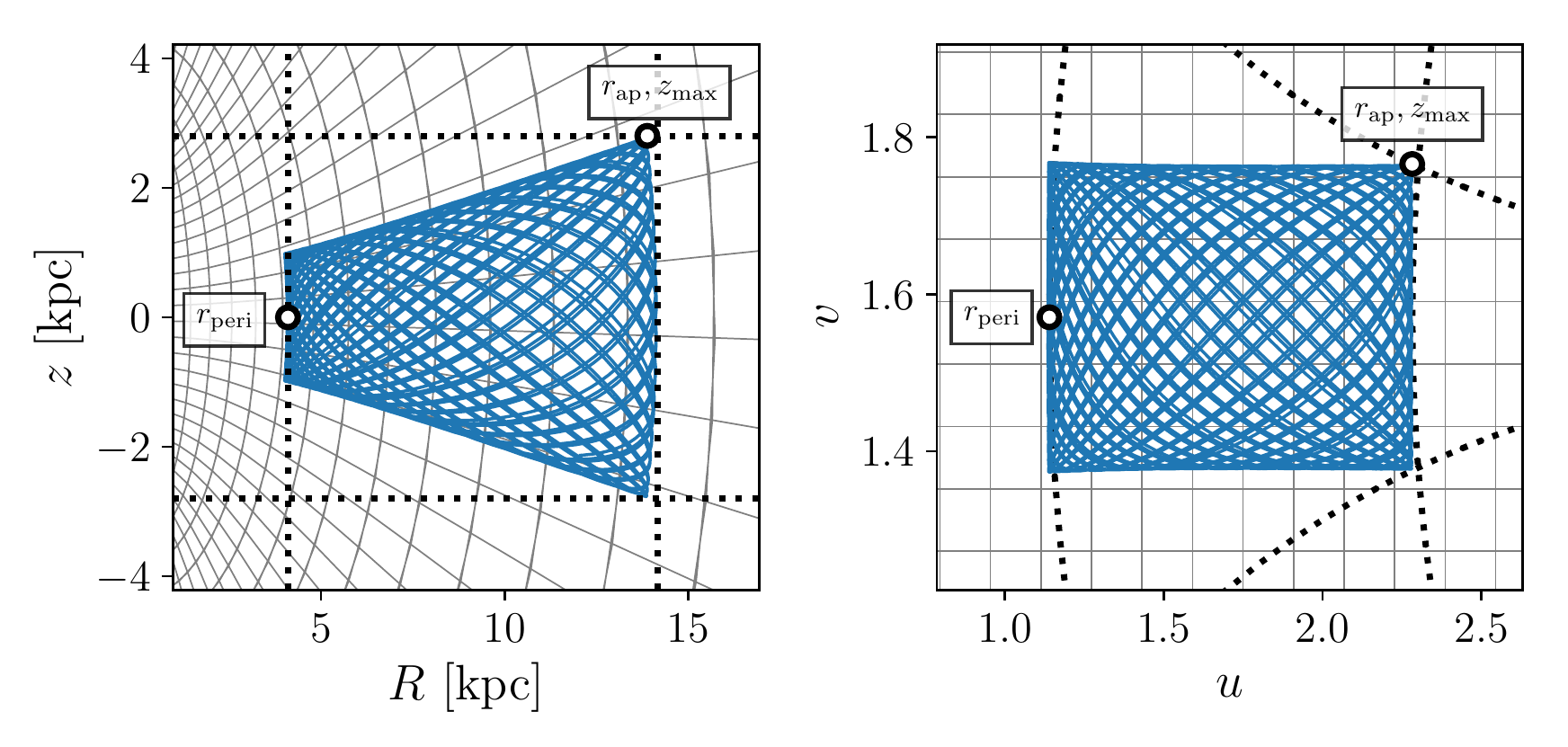}
\caption{\label{fig:example-orbit}An example orbit in \texttt{MWPotential2014}. The left panel shows the orbit as integrated in the $R-z$ plane, whereas the right panel demonstrates the same orbit under transformation into the $u-v$ plane. A grid of constant $u-v$ is shown in both panels in grey. The large dots in both panels demonstrate the point in the orbit where $r_{\mathrm{peri}}$, $r_{\mathrm{ap}}$ and $z_{\mathrm{max}}$ are reached. The vertical (horizontal) dotted lines in each panel show the locations of $R=r_{\mathrm{peri}}, R=r_{\mathrm{ap}}$ (and $z=z_{\mathrm{max}}$). The orbit projected onto the $u-v$ plane is rectangular to an excellent approximation.}
\end{centering}
\end{figure*}

Many galactic mass distributions, and in particular that of the Milky Way, are well approximated by a St\"ackel potential \citep[e.g.,][]{1985MNRAS.216..273D,1988ApJ...329..720D}. These potentials are defined in terms of prolate confocal coordinates (see \citealt{2008gady.book.....B}), with
\begin{align}\label{eq:prolatetransform1}
R & = \Delta \sinh u \sin v\\
z & =\Delta \cosh u \cos v\,,\label{eq:prolatetransform2}
\end{align}
where $\Delta$ is a parameter that specifies the focal point of the coordinate system, placed at $R=0$, $z=\pm\Delta$. The momenta in these coordinates are then given by
\begin{equation}
\begin{aligned}
p_u & = \Delta (p_R \cosh u \sin v + p_z \sinh u \cos v)\\ 
p_v & = \Delta (p_R \sinh u \cos v - p_z \cosh u \sin v)\,.
\end{aligned}
\end{equation}
In these coordinates, an oblate, axisymmetric St\"ackel potential is a potential that can be written in terms of two functions $U(u)$ and $V(v)$ of one variable as 
\begin{equation}\label{eq:staeckelpot}
\Phi_\mathrm{S}(u,v) = \frac{U(u)-V(v)}{\sinh^2u+\sin^2v}\,.
\end{equation}
For a potential of this form, the Hamilton-Jacobi equation can be solved using the separation-of-variables method, the motions in $u$ and $v$ decouple, and we have that
\begin{equation}
\begin{aligned}
\frac{p_{u}^2}{2\Delta^2} & = E \sinh^{2}u-I_3-U(u)-\frac{L_z^2}{2\Delta^2\sinh^2u} \\
\frac{p_{v}^2}{2\Delta^2} & = E \sin^{2}v-I_3-V(v)-\frac{L_z^2}{2\Delta^2\sin^2v}
\end{aligned}
\end{equation}
where $E$ and $L_z$ denote the energy and vertical component of the
angular momentum of the orbit, respectively, and $I_3$ is a constant
of separation---the third integral. By numerically solving the
equations for the turning points, $p_u(u) = 0$ and $p_v(v) = 0$, one
can determine the spatial boundary of the orbit, which is rectangular
in $(u,v)$: $u_\mathrm{min/max}$ in $u$ and $v_{\mathrm{min/max}}$ in
$v$. For a St\"ackel potential that is symmetric around the midplane,
$v_\mathrm{min} = \pi-v_\mathrm{max}$ and we will assume that this
holds hereafter.

For galactic potentials $\Phi$ that are close to a St\"ackel potential
but not exactly equal to one, Equation \eqref{eq:staeckelpot} only
approximately holds. Following \citet{2012MNRAS.426.1324B}, for such
potentials we can define functions $U(u)$ and $V(v)$ as
\begin{align}
  U(u) & \equiv \cosh^2 u\, \Phi(u,\pi/2)\\
  V(v) & \equiv \cosh^2 u_0\, \Phi(u_0,\pi/2) - (\sinh^2 u_0+\sin^2 v)\,\Phi(u_0,v)\,,
\end{align}
using a reference point $(u,v) = (u_0,\pi/2)$, to create an approximate
St\"ackel potential $\Phi_S$ for $\Phi$ using Equation
\eqref{eq:staeckelpot}. Using these functions, we can solve for the
spatial boundary of the orbit in $\Phi_S$, which approximates the
boundary in the desired potential $\Phi$. When computing the boundary
for a single phase-space point we simply set $u_0$ to the $u$
coordinate of the phase-space point; when we generate an interpolation
grid as described below, we use the method described by
\citet{2012MNRAS.426.1324B} to determine a good $u_0$ as a function of
$E$ and $L_z$.

By transforming the rectangular boundary back to $(R,z)$ using Equations \eqref{eq:prolatetransform1} and \eqref{eq:prolatetransform2}, we can determine the orbit's spatial boundary in regular cylindrical coordinates. This boundary is typically summarized using the peri- and apogalacticon distances $r_{\mathrm{peri/ap}}$---which we take to be the closest and furthest three-dimensional distance to the galactic centre along the orbit---and the maximum height $z_{\mathrm{max}}$ above the plane. From the geometry of the prolate spheroidal coordinate system, it is straightforward to see that the perigalacticon is attained at $z=0$ ($v = \pi/2$) and $u = u_\mathrm{min}$
\begin{equation}
  r_\mathrm{peri} = \Delta \sinh{u_{\mathrm{min}}} \,,
\end{equation}
while $z_\mathrm{max}$ and the apogalacticon are both attained at $(u,v) = (u_\mathrm{max},v_\mathrm{max})$
\begin{equation}
\begin{aligned}
z_{\mathrm{max}} & = \Delta \cosh u_{\mathrm{max}} \cos v_\mathrm{min} \\
r_{\mathrm{ap}} & = \sqrt{(\Delta \sinh{u_{\mathrm{max}}} \sin{v_\mathrm{min}})^2+z_{\mathrm{max}}^2}\,.
\end{aligned}
\end{equation}
We can then also compute the orbital eccentricity using its usual definition for galactic orbits
\begin{equation}\label{eq:eccentricity}
e = \frac{(r_{\mathrm{ap}}-r_{\mathrm{peri}})}{(r_{\mathrm{ap}}+r_{\mathrm{peri}})}.
\end{equation}

This computation can be performed relatively quickly and inexpensively
and is straightforward to parallelise for large numbers of orbits. A
good value for the parameter $\Delta$, which defines the prolate
coordinate system, can be computed for a given $(R,z)$ position using
Equation (9) of \citet{2012MNRAS.426..128S}, which exploits the
relation between $\Delta$ and the first and second derivatives of the
potential that holds for a St\"ackel potential to determine a good
$\Delta$ using the derivatives of any axisymmetric potential.

We can further speed up the computation of the orbital parameters
$(e,z_\mathrm{max},r_\mathrm{peri},r_\mathrm{ap})$ by building an
interpolation grid using the implementation described in Section 5.4
of \citet{2015ApJS..216...29B} of the interpolation method first
discussed in \citet{2012MNRAS.426.1324B}.

In Fig. \ref{fig:example-orbit}, we demonstrate the appearance of an exemplar orbit in the \texttt{MWPotential2014} Milky-Way-like potential described in \citet{2015ApJS..216...29B}, in both cylindrical $R,z$ coordinates and under the transformation into the $u,v$ plane described above. The orbit shown is at a random energy equivalent to $\log_{10}\left(\frac{E-E_c(L)}{E(\infty)-E_c(L)}\right) = -0.8$, and an angular momentum $\log_{10}(L) = -0.1$ in units of the angular momentum of the circular orbit at the Sun. The orbit is squashed from a cone-like shape in $R-z$ to an approximate box-like geometry in the $u-v$ plane. In this geometry the vertical and radial oscillation is readily separable, allowing the simple calculation of the parameters.  

\subsection{Implementation in \texttt{galpy}}

\begin{figure*}
\begin{centering}
\includegraphics[width=0.75\textwidth]{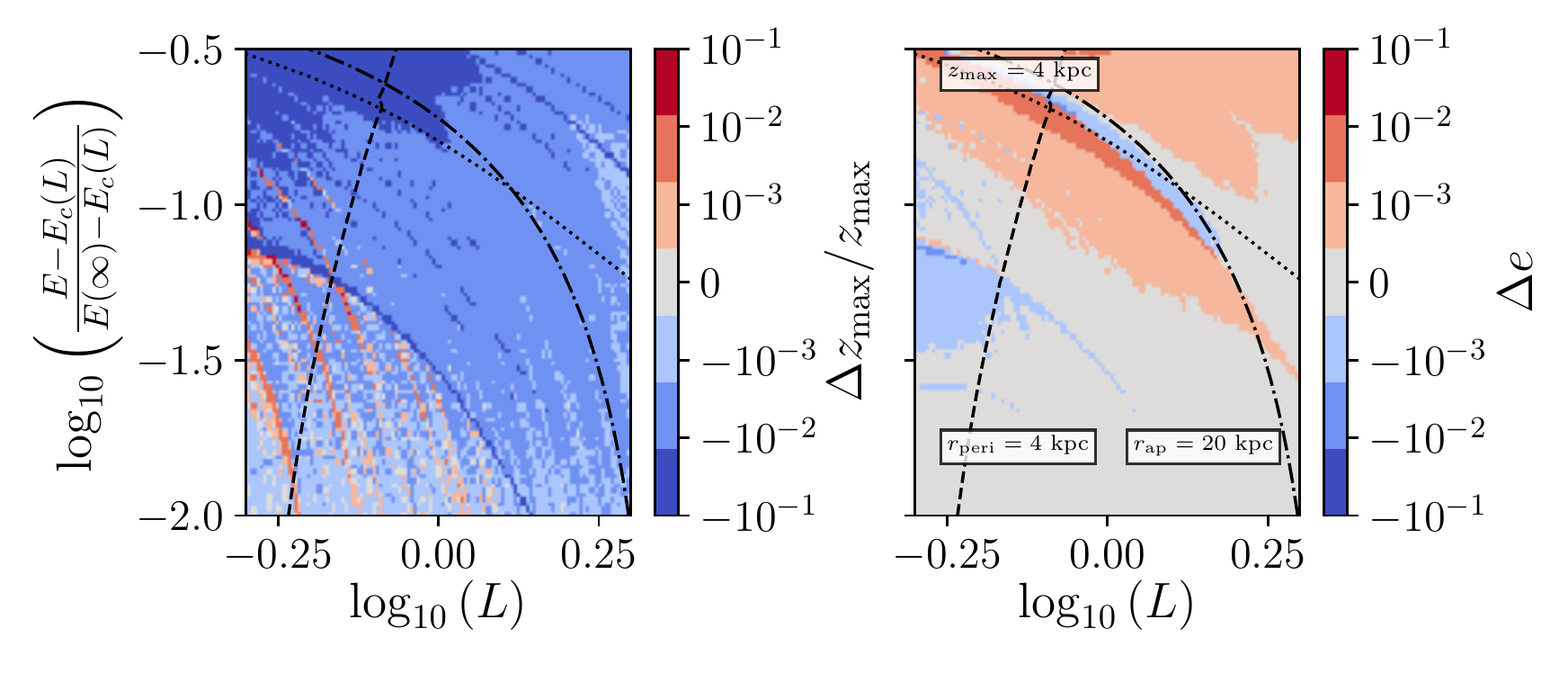}
\caption{\label{fig:diskcomp} The difference between orbital parameter estimation in the Milky-Way-like potential \texttt{MWPotential2014} by direct orbit integration and using the St\"ackel approximation from Section 2 for a set of orbits with angular momenta $-0.3 < \log_{10}(L) < 0.3$ and random energies $-2 < \log_{10}\left(\frac{E-E_c(L)}{E(\infty)-E_c(L)}\right) < -0.5$. The left panel shows the difference between integrated and estimated $z_\mathrm{max}$ normalised to the integrated $z_\mathrm{max}$ value, whereas the right panel shows the difference between integrated and estimated $e$, without normalisation. The dashed, dash-dotted and dotted lines follow the energy and angular momenta values where $r_\mathrm{peri}=4$ kpc, $r_\mathrm{ap}=20$ kpc and $z_\mathrm{max}=4$ kpc, respectively. Inside this region, both parameters are estimated to a very high level of accuracy when compared to the orbit integration. There are sharp regions where the estimation is not as accurate, which are related with regions of energy and angular momentum space where the orbit tori are not well filled by orbit integration.}
\end{centering}
\end{figure*}

We have implemented the method described above in the \texttt{galpy} galactic dynamics Python
package\footnote{\url{https://github.com/jobovy/galpy}}
\citep{2015ApJS..216...29B} and included it in its recent \texttt{v1.3} release, both in its direct form and in the grid-based form. In this way, the method can be used for \emph{any} orbit in \emph{any} axisymmetric potential that is implemented in \texttt{galpy}. We briefly describe the details of this implementation here.

The novel method for determining orbital parameters described here is naturally a part of the St\"{a}ckel approximation for computing orbital actions and angles, which is implemented in \texttt{galpy}  as a class \texttt{actionAngleStaeckel} with methods that return actions, frequencies, and angles. We therefore implemented a new method \texttt{EccZmacRperiRap} of these objects that uses the formalism above to compute the orbital eccentricity $e$, maximum vertical excursion $z_{\mathrm{max}}$, and peri- and apogalacticon radii $r_\mathrm{peri}$ and $r_\mathrm{ap}$. The \texttt{EccZmacRperiRap} method uses a C implementation of the method if the provided gravitational potential has a C implementation---which is the case for almost all built-in potentials---and falls back onto a pure Python implementation otherwise. The \texttt{EccZmacRperiRap} method can be applied to arrays of phase-space positions and can use a different $\Delta$ parameter for each phase-space position. The C implementation can furthermore make use of OpenMP to parallelise the calculation for different phase-space positions. The grid-based method is implemented by adding a method \texttt{EccZmacRperiRap} to the \texttt{actionAngleStaeckelGrid} class in \texttt{galpy}---which implements the grid-based version of the algorithm of \cite{2012MNRAS.426.1324B}---that uses a grid of $e$, $z_{\mathrm{max}}$, $r_\mathrm{peri}$, and $r_\mathrm{ap}$ pre-computed during the instantiation of a \texttt{actionAngleStaeckelGrid} object. The interpolation is performed in the same way as that of the actions (see \citealt{2015ApJS..216...29B} for details on this).

The interface through the \texttt{actionAngleStaeckel} class allows large numbers of phase-space points to be processed quickly, but requires the phase-space points to be input in Galactocentric cylindrical coordinates and a $\Delta$ parameter (or array of such parameters) to be given. A simpler interface to the same method is provided through \texttt{galpy}'s \texttt{Orbit} class, which represents galactic orbits and forms the basis of orbit integration in \texttt{galpy}. \texttt{Orbit} instances can be initialized in a variety of ways, including from observed positions and velocities (sky coordinates, distances, proper motions, and line-of-sight velocities). The fast method of this paper is implemented as part of the existing \texttt{e}, \texttt{zmax}, \texttt{rperi}, and \texttt{rap} methods of \texttt{Orbit} instances. This interface performs an automatic determination of a good $\Delta$ parameter (using
Equation (9) of \citealt{2012MNRAS.426..128S} applied to the current position). Moreover, for spherical potentials $\Delta = 0$ and the \texttt{Orbit} methods automatically detect this and use the simpler version of the method above that is appropriate for spherical potentials.

We provide some explicit code examples in Sec. \ref{sec:use_python} below.

\section{Tests and Applications}

\begin{figure*}
\begin{centering}
\includegraphics[width=0.95\textwidth]{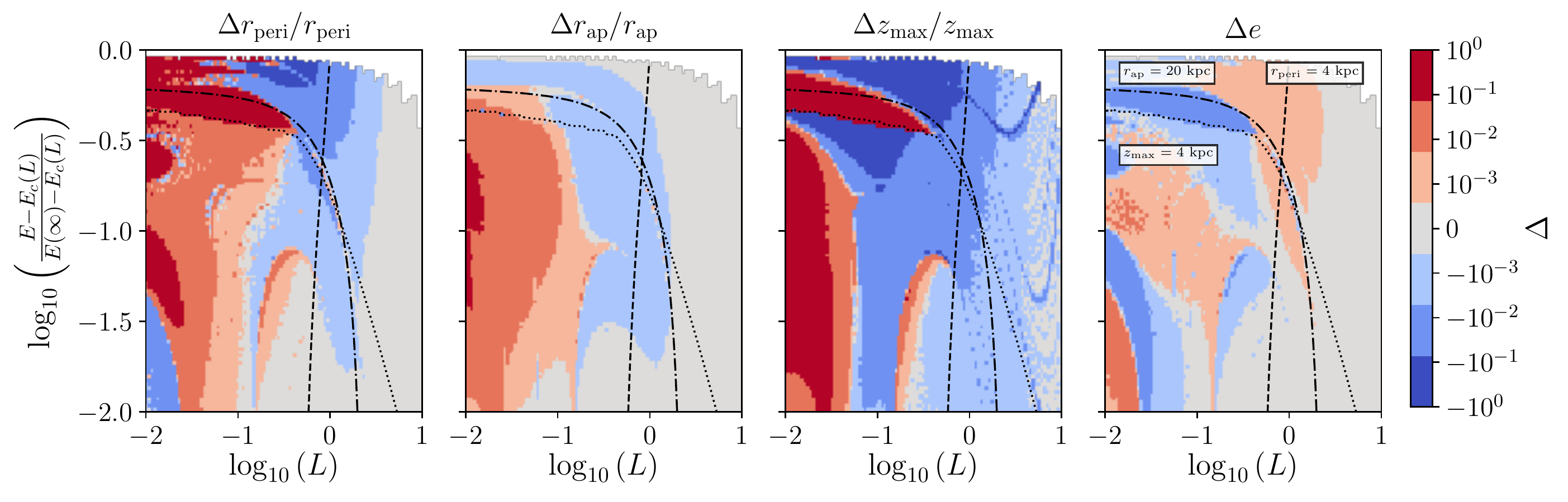}
\caption{\label{fig:comparison} An equivalent comparison to that shown in Fig. \ref{fig:diskcomp} with an extended range in random energy and angular momentum (corresponding to guiding radii ranging from $\sim 0.2$ to 117 kpc). The comparison between integration and estimation is now also shown for $r_\mathrm{peri}$ and $r_\mathrm{ap}$ in the top two panels and, as above, these are normalised to the value of the parameter from orbit integration. The method is still highly accurate across a wide range of relevant orbits, with only isolated regions showing large deviations from the integration value. These regions are found to be occupied by resonant orbits in most cases, where the estimation is not valid.}
\end{centering}
\end{figure*}

In the following, we demonstrate the accuracy of the estimation of orbit parameters via the methodology described in Section \ref{sec:method}. We use the parameters calculated using an orbit integration technique as `truth' values in each case, but note that these calculations are subject to some uncertainty, arising from the (small) error in the integration, and in the subsequent calculation of parameters such as the eccentricity, which can be underestimated if the orbit torus is not fully filled by the integration.

\subsection{Estimating parameters of disc orbits}
\label{sec:disc_orbits}

First, we demonstrate the accuracy of estimation of $z_{\mathrm{max}}$ and $e$ in a grid of orbits in \texttt{MWPotential2014} spanning the range of angular momentum $-0.3 < \log_{10}(L) < 0.3$ (angular momentum here and everywhere below is expressed in units of the angular momentum of the circular orbit at the Sun) and covering at each $L$ the range of random energy $-2 < \log_{10}\left(\frac{E-E_c(L)}{E(\infty)-E_c(L)}\right) < -0.5$. This region of orbital space roughly corresponds to that of stars on disk orbits, with angular momenta corresponding to guiding radii $3 \lesssim R_{\mathrm{g}} \lesssim 30\ \mathrm{kpc}$. At each energy and angular momentum point, we integrate an orbit for 20 azimuthal periods (at fixed timestep), initialised at $R=R_{\mathrm{g}}$ and $z=0$ kpc, with a tangential velocity $v_T = L/R_{\mathrm{g}}$ and radial and vertical velocity $v_R$ and $v_z$, such that 
\begin{equation}
\label{eq:xvel}
v_R = \sqrt{x [E-E_c(L)]} \\
v_z = \sqrt{(1-x)[E-E_c(L)]},
\end{equation} where, here, we let $x=4/5$. We find that integration for 20 azimuthal periods is sufficient to estimate the orbital parameters with a precision better than a hundredth of a percent for the orbits shown. The `true' parameters of the orbit are then calculated based on this integration. We then estimate the $\Delta$ parameter required for the application of the method using the method from \citet{2012MNRAS.426..128S} as described above, taking the median estimated value for a range of phase space points along a small part of the orbit. Using this, we estimate the parameters again using the St\"ackel approximation method, then compare these with the integrated value. We show the results in Fig. \ref{fig:diskcomp} by plotting the difference between the integrated parameter and estimated parameter, $\Delta\ \mathrm{P} = \mathrm{P}_{\mathrm{integrated}}-\mathrm{P}_{\mathrm{estimation}}$, where P represents the parameter in question and where, in the case of $z_{\mathrm{max}}$, we normalise this value by the integrated parameter.
The values of energy and angular momenta where $r_{\mathrm{peri}}=4\ \mathrm{kpc}$, $r_{\mathrm{ap}} = 20\ \mathrm{kpc}$, and $z_{\mathrm{max}} =4\ \mathrm{kpc}$ are indicated by \emph{dashed, dash-dotted} and \emph{dotted} lines, respectively.

For both of the parameters, $z_{\mathrm{max}}$ and $e$, the median difference between the estimated and integrated parameter is much less than 1\%. For the majority of orbits in the region bounded by the lines of constant $r_{\mathrm{peri}},r_{\mathrm{ap}}$ and $z_{\mathrm{max}}$ and centered on $\log_{10}(L) = 0$ (hereafter referred to as the disc region), the parameters are extremely well estimated. The left panel, showing $\Delta z_{\mathrm{max}}/z_{\mathrm{max}}$, demonstrates that there is a small, roughly systematic offset between integration and estimation, at a level of $\sim 10^{-3}$. A number of orbits are more strongly overestimated relative to the integration value, with significant, sharp substructure in energy-angular momentum space. %{\bf TED: In the rest of this paragraph, do you mean |rel. diff.|? I think you do, so would be good to explicitly say $|\Delta e|$ and similar to make this clear. If that's not what you mean, I think we want absolute values (because you already described the bias).}
However, the median $|\Delta z_{\mathrm{max}}/z_{\mathrm{max}}|$ across all the orbits shown is still small, at $\sim 4\times10^{-3}$. We demonstrate in the right panel, showing $\Delta e$, that the estimation is accurate to a level less than $0.1\%$ in the disc region, with no obvious systematic offset relative to the integration. The median $|\Delta e|$ over the full range of energy and angular momentum shown is small, at $\sim 6\times10^{-5}$. Although there are regions of the space where the parameters are over- and underestimated, these are localised to the sharp regions, which appear to correspond to areas in the energy-angular momentum space where the orbital tori are not well filled by the orbit integration.

We also repeat this test across a grid in the $x$ parameter in Equation \eqref{eq:xvel}, performing the estimation in the same random energy and angular momentum grid, but varying $x$ between 0 and 1. We find that this initial ratio between the radial and vertical velocities makes little difference to the performance of the estimation, and that $\Delta \mathrm{P}$ correlates more strongly with the energy and angular momentum. We note, however that the performance of the estimation is slightly worsened, to the order of a few tenths of a percent in all the parameters, at the edges of the $x$ grid, where the random velocity is concentrated almost entirely radially or vertically.

\subsection{Applying the method to a wider range of orbits}

We now extend the range of energy and angular momenta to look at orbits in with $-2 < \log_{10}(L) < 1$ (with guiding radii ranging from $\sim 0.2$ to 117 kpc) and $-2 < \log_{10}\left(\frac{E-E_c(L)}{E(\infty)-E_c(L)}\right) < 0$. We study the accuracy of estimation in $r_{\mathrm{peri}}$ and $r_\mathrm{ap}$, as well as the parameters shown in the previous section. We follow the same procedure as described in Section \ref{sec:disc_orbits}, showing the results in Fig. \ref{fig:comparison}, retaining the definition of $\Delta \mathrm{P}$, and choosing to also normalise this value by the integrated parameter in the cases of $r_{\mathrm{peri}}$ and $r_\mathrm{ap}$.

In this wider range of orbits, there are clearly cases where the estimation has difficulty matching the calculation returned by the orbit integration, with some parameters being returned at values greater than $\sim 60\%$ different in isolated regions of energy and angular momentum space. The method particularly breaks down at lower angular momenta ($\log_{10}(L)\lesssim -1$ or $R_g \lesssim 1.17\ \mathrm{kpc}$). Direct inspection of the orbits in these regions shows that they are generally occupied by resonant orbits, which do not fill the orbital tori which are described by the St\"ackel approximation. It is also worth noting that the St\"ackel approximation breaks down in the regions of parameter space where the potential deviates strongly from a St\"ackel potential. Regardless, across the whole range of orbits, the method still achieves median variations of order $10^{-3}$ over all the parameters, demonstrating the isolation of the regions where it is not so accurate.

\subsection{Using different potentials}

All tests thus far have used \texttt{MWPotential2014}. The components of this potential, and the dynamical constraints used to fit their parameters are described in \citet{2015ApJS..216...29B}. Here, we test an alternate and more complex Milky Way-like potential to understand how the assumed potential might affect the estimation. We choose to implement the best-fitting Milky Way potential of \citet{2017MNRAS.465...76M}, consisting of a flattened axisymmetric bulge, an NFW halo, pure exponential thin and thick disks, and two gas discs, representing the \ion{H}{1} and molecular gas, with $\sech^2$ vertical profiles and exponential radial profile (including a hole in the centre with an exponential scale length). We approximate the bulge and the four discs using \texttt{galpy}'s \texttt{SCFPotential} and \texttt{DiskSCFPotential}, respectively. These are implementations of the Self-Consistent Field (SCF; \citealt{1992ApJ...386..375H}) method for generating potentials from general density functions; \texttt{DiskSCFPotential} uses the approach of \citet{1995MNRAS.277.1341K} to approximate the disc contribution to a potential and solves for the (approximately spherical) difference between the approximate and true potential using the SCF method. Our implementation approximates the \citet{2017MNRAS.465...76M} density to better than $1\,\%$ everywhere. We numerically compute the second derivatives of the potential for the estimation of the St\"ackel $\Delta$ parameter.

We repeat the comparison between orbit integration and the estimation method as in Section \ref{sec:disc_orbits} using the alternative potential. We find a median $|\Delta z_{\mathrm{max}}/z_{\mathrm{max}}| = 0.034$, in a region of energy and angular momentum equivalent to that shown in Fig. \ref{fig:diskcomp}, which also contains a region bounded by orbit parameters consistent a `disc' orbit in this potential. We find a median $|\Delta e| = 3\times 10^{-5}$ in the same region. The estimation of $z_\mathrm{max}$ in this potential is significantly worse than that in \texttt{MWPotential2014}, but $e$ is still estimated to a high level of accuracy. The worse performance for $z_\mathrm{max}$ is due to the fact that the vertical structure of this potential is more complex than that of \texttt{MWPotential2014} and the potential is therefore less well approximated as a St\"ackel potential. The structure in the energy-angular momentum plane as shown in Figures \ref{fig:diskcomp} and \ref{fig:comparison} is still present, with most of the differences between the potentials being systematic in nature.

We also compare the estimation of parameters in this potential with the same in \texttt{MWPotential2014}. We find that there are significant systematic offsets between the two sets of results. The disc region, which is similar between the two potentials has a median difference in $z_\mathrm{max}$ of $\sim 15\%$, and in $e$ of $\sim 5\%$. The offset between potentials is roughly systematic in $z_\mathrm{max}$, whereas those in $e$ vary as a function of angular momentum. 

\subsection{Validating the method with a real dataset}

\begin{figure}
\begin{centering}
\includegraphics[width=0.9\columnwidth]{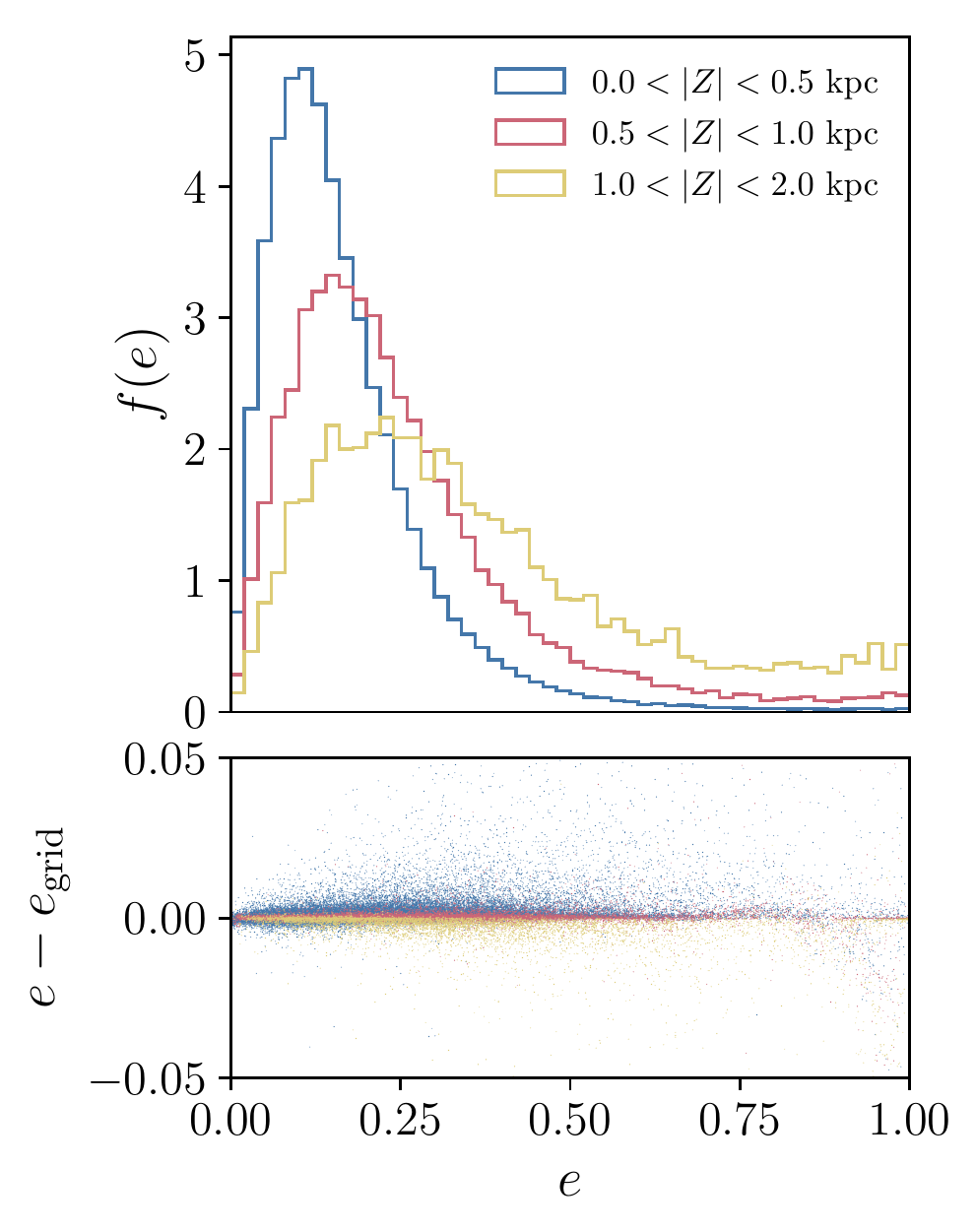}
\caption{\label{fig:rave-gaia-e} The eccentricity distribution of the RAVE-\emph{Gaia}-TGAS data set as a function of height above the midplane $|Z|$, as calculated using the St\"ackel approximation method. The eccentricity distribution becomes broader, and peaks at higher $e$ at greater $|Z|$. The bottom panel shows a comparison between eccentricity values as calculated serially $e$, and those calculated via a grid-based interpolation method $e_\mathrm{grid}$.}
\end{centering}
\end{figure}

To demonstrate the validity of the method in comparison to orbit integration, we apply it to the cross section of the RAVE and \emph{Gaia}-TGAS datasets; a sample of 216,201 stars with position, distance, proper motion and heliocentric line-of-sight velocity measurements. This data set is useful for validating the method, as it is possible to (relatively) quickly measure the parameters via orbit integration as a cross-check.  We cross match the two data sets by sky position, then convert the observed coordinates into Galactocentric cylindrical coordinates, assuming the solar radius and height above the midplane $R_0=8\ \mathrm{kpc},Z_0 = 0.025\ \mathrm{kpc}$ and a circular velocity $V_c = 220\ \mathrm{km\ s^{-1}}$ \citep[e.g.][]{2012ApJ...759..131B}. We assume the solar motion of \citet{2010MNRAS.403.1829S}: $[U,V,W]_{\odot} = [-11.1,12.24,7.25]\ \mathrm{km\ s^{-1}}$. We then estimate the orbital eccentricities in \texttt{MWPotential2014} using the direct method, estimating $\Delta$ at each phase-space point, and also applying the grid-based method with a fixed $\Delta=0.4\,R_0$. We find that the regular method returns values for the entire sample in $\sim 20$ s, whereas the grid-based method performs the same estimation in $\sim 2$ s. The speed of the estimation depends on the complexity of the potential. Under a simpler, flattened logarithmic halo potential results are returned in $\sim 400$ ms. 

We show the estimated eccentricity distribution across three bins in vertical distance from the midplane $|Z|$, in the upper panel of Fig. \ref{fig:rave-gaia-e}. The lower panel shows a comparison between the regular and grid method estimation values. We find that the median eccentricity increases with $|Z|$, and that the distribution becomes broader, such that the relative fraction of stars on eccentric orbits becomes larger as $|Z|$ increases, in good agreement with existing measurements \citep[e.g.][]{2010ApJ...725L.186D,2013A&A...554A..44A,2015MNRAS.447.3526K}. The grid-based method generally agrees very well with the regular method, with a standard deviation in $e-e_{\mathrm{grid}}$ less than $\sim 10^{-2}$. It is noteworthy that the grid estimation returns values marginally closer to the regular method ($\lesssim 0.5\%$ more accurate) for stars at intermediate $|Z|$, with the smallest and largest $|Z|$ bins returning values with similar accuracy. It should be noted that stars which are in regions of angular momentum-energy space which are not well approximated by the St\"ackel fudge can be subject to systematic offsets in $e$ up to $\sim 10^{-2}$. These offsets may act to artificially broaden the eccentricity distributions, but the observed broadening as a function of $|Z|$ exceeds these uncertainties.

\subsection{A Catalogue of estimated orbital characteristics for the \emph{Gaia} DR2 RV sample}

\begin{figure}
\includegraphics[width=\columnwidth]{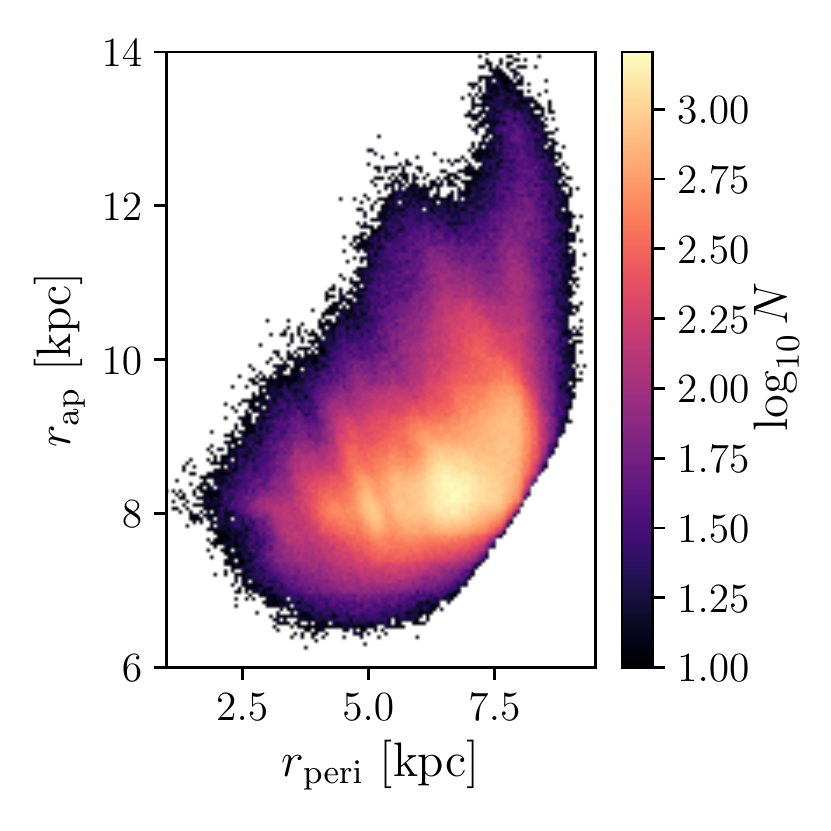}
\caption{\label{fig:rperirap} The distribution of stars in the \emph{Gaia} DR2 RVS sample, with $\varpi/\delta \varpi > 20$, in $r_\mathrm{ap}$ and $r_\mathrm{peri}$. Bins with $< 10$ stars are not coloured. There are many substructures in this plane, the understanding of which will likely be of great importance to the understanding of the formation and evolution of the Milky Way. However, for the purpose of this work, it demonstrates that while spatially limited, \emph{Gaia} DR2 samples a wide range of Galactic orbits, that range from the center of the Galaxy to far out into the Galactic Halo.}
\end{figure}

We now demonstrate the usefulness of this method and the previously examined St\"ackel approximation for action-angle coordinates of \citet{2012MNRAS.426.1324B} in estimating orbital characteristics for a sample of a size that would be computationally intractable using orbit integration. We release with this paper a catalogue of estimated orbit parameters: $e, r_\mathrm{ap}, r_\mathrm{peri}, Z_\mathrm{max}$, action-angle coordinates: $J_R, L_z, J_z, \theta_R, \theta_\phi, \theta_z$, orbital frequencies: $\Omega_R, \Omega_\phi, \Omega_z$, orbit energies and guiding radii for the \emph{Gaia} DR2 stars with a 6D phase-space solution (parallax, celestial position, proper motion and radial velocity). The catalogue assumes a left-handed coordinate frame (positive solar angular momentum). The angles are approximated such that the radial angle $\theta_R$ is zero at pericenter, increasing towards the apocenter, the vertical angle $\theta_z$ is zero at $z=0$, increasing toward positive $z_\mathrm{max}$.

We take the full \texttt{gaia\_source\_with\_rvs} table from the \emph{Gaia} Archive, and for each object with a measurement of all the necessary parameters (i.e. no NULL fields, a total of 6643147 stars), we sample 100 realisations of the observation by reconstruction of the covariance matrix of the astrometric parameters, which are given in the table. The orbital characteristics listed above are then estimated for each realisation, given the simple \texttt{MWPotential2014}, and the mean and standard deviation reported for the parameter value and its associated error. We also compute and report the correlation coefficients between the orbital parameters and the actions. This process is obviously more time consuming than a simple operation of the method over a list of 6D coordinates, as it requires the sampling of the covariance matrix, and the computation of the correlation coefficients for 6643147 stars. Estimation of the uncertainties by this method also means that we effectively perform 100 times more estimations. In total, the estimations that form the catalogue take $\sim 116$ hours, parallelised across 16 cores. Performing a similar computation using orbit integration would obviously take considerably longer.

The table is available in the supplementary material which accompanies the online article. We use the data model provided in Table \ref{tab:datamodel}. The provided table is ordered by the \emph{Gaia} integer source ID, and contains rows populated with \texttt{NaN} values for objects that did not have the necessary astrometric parameters, so that it can be directly joined row-for-row to a flattened \texttt{gaia\_source\_with\_rv} file. The table also includes \emph{Gaia} source ID's, and so is readily joinable with the tables available on the Gaia archive.

As a simple demonstration of the scientific value of the catalogue as we provide it, in Figure \ref{fig:rperirap} we show the density of stars in $r_\mathrm{ap}$-$r_\mathrm{peri}$ space, for stars with parallax signal-to-noise ratio $> 20$. While the appearance of this plane is clearly affected by the selection function of the higher quality part of the $\emph{Gaia}$ RVs sample, which is restricted mainly to regions close to the Sun, it shows that the \emph{orbits} sampled by \emph{Gaia} extend from the very inner galaxy, well into the halo. This is a clear testament to the scientific value of this dataset. There are also substructures visible in this plane, the understanding of which will provide detailed insight into the orbital structure of the Galaxy.

\begin{figure*}
\includegraphics[width=\textwidth]{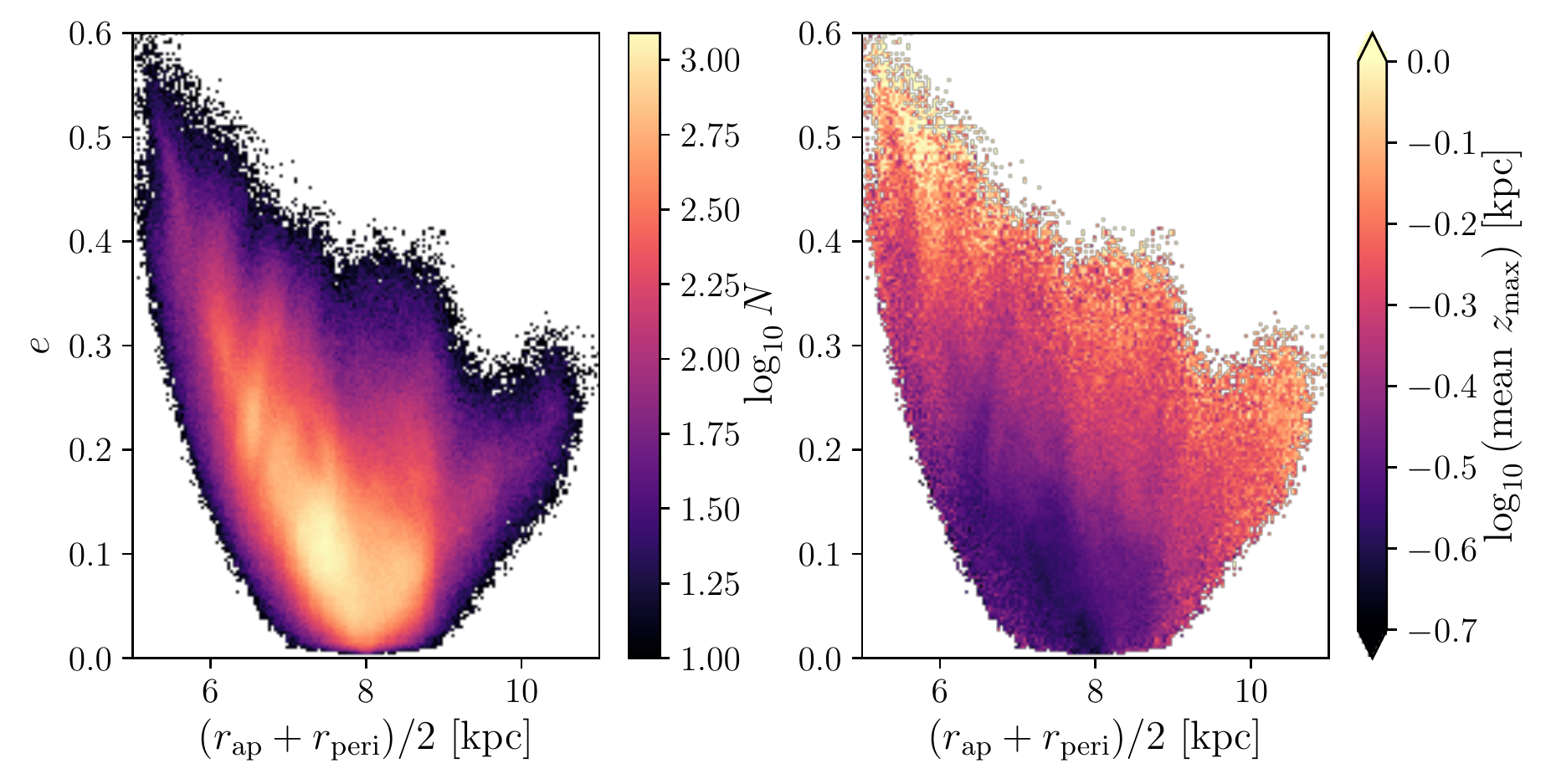}
\caption{\label{fig:rmeane} The distribution of stars in a local ($1/\varpi < 1.5$ kpc) \emph{Gaia} DR2 RVS sample in $r_\mathrm{mean}$ and $e$ space, where $r_\mathrm{mean} = (r_\mathrm{ap}+r_\mathrm{peri})/2$. In the left panel, we show the density of stars in the space, where many sub-structures are visible, which are at least qualitatively similar to those seen in action space. In the right hand panel, we demonstrate that these overdensities in $r_\mathrm{mean}$-$e$ correspond to regions with lower mean $z_\mathrm{max}$, suggesting that they may have arised due to resonances from `disk' dynamics related to, for example, bars and spiral features, rather than e.g. `halo' dynamics such as satellite flybys or subhalo bombardment.}
\end{figure*}

A simple exploration of the orbit space structures in Figure \ref{fig:rperirap}, which may provide insight into their origin, is shown in Figure \ref{fig:rmeane}. The left hand figure shows the density of stars in the space of $(r_\mathrm{ap}+r_\mathrm{peri})/2 = r_\mathrm{mean}$ vs. $e$, and the mean $z_\mathrm{max}$ in the same bins on the right. This plane is roughly analogous to the $J_R$-$L_z$ plane in action space \citep[explored in \emph{Gaia} DR2 by][]{2018arXiv180503653T}, but offers a more direct and intuitive link to the spatial and kinematic structure of the disk. As in action-space, and in our Figure \ref{fig:rperirap}, there are many clumpy features in this space, which suggest that the orbital distribution function is not smooth, but rather stars are `trapped' in certain regions of the space. 

In the right hand panel, we show that the clumpy features in  $r_\mathrm{mean}$-$e$ space correspond to regions where the mean $z_\mathrm{max}$ is low. This suggests that these features are more likely to be induced by disk dynamical effects which act in-plane, rather than halo dynamics, such as satellite flybys and bombardment by small subhaloes. This is consistent with the results of \citet{2018arXiv180503653T}, who showed that the clumps in action space had a low vertical action, $J_z$ (which is analogous to $z_\mathrm{max}$). Through proper modelling of this plane in orbital parameter and action space, it will be possible to begin linking these features to the complex dynamics of the Milky Way, which will give way to new insights into its present day structure and history of its formation and evolution. 

In terms of uncertainties on the estimated parameters, and their implications as to the conclusions drawn from the catalogue, there are three main sources of uncertainty to consider: observational errors, choice of potential, and systematics from the application of the estimation method. In general, uncertainties arising from observational errors are small, and of the order of a few percent. We have shown in the earlier sections of this paper that systematic uncertainties arising from the use of the estimation method are also small, at a level less than a percent over wide ranges of energy-angular momentum space. Most importantly, we have discussed that the choice of potential imposes significant systematic uncertainties on these estimates. Adopting a more complex potential such as that of \citet{2017MNRAS.465...76M} can change these estimates systematically by as much as $20\%$, and so this should be taken into account when interpreting the tabulated parameters. 

\begin{deluxetable*}{lccc}
\caption{\label{tab:datamodel} The adopted data model in the table of estimated orbital parameters, actions, frequencies, angles, energies and guiding radii for stars in the \emph{Gaia} DR2 RVS catalogue. For brevity, we denote the repeated error and correlation columns as \texttt{*\_err} and \texttt{*\_*\_corr}, where \texttt{*} should be replaced by the given column ID for the desired quantity(ies).}
\tablehead{Column ID & Quantity & Description & Units}
\startdata
\texttt{source\_id} & \emph{Gaia} DR2 source ID & Identifier provided in DR2 & \\
\texttt{ra} & R.A. & the object right ascension & Deg. \\
\texttt{dec} & Dec & the object declination & Deg. \\
\texttt{e} & $e$ & orbital eccentricity as defined in Equation \eqref{eq:eccentricity} & \\
\texttt{z\_max} & $z_\mathrm{max}$ & maximum vertical excursion from the midplane & kpc \\
\texttt{r\_peri} & $r_\mathrm{peri}$ & 3D pericenter radius & kpc\\
\texttt{r\_ap} & $r_\mathrm{ap}$ & 3D apocenter radius & kpc \\
\texttt{jr} & $J_R$ & radial action & $\mathrm{km\ s^{-1}\ kpc}$ \\
\texttt{Lz} & $L_z$ ($J_\phi$) & azimuthal action (equivalent to vertical component of angular momentum) & $\mathrm{km\ s^{-1}\ kpc}$ \\
\texttt{jz} & $J_Z$ & vertical action & $\mathrm{km\ s^{-1}\ kpc}$ \\
\texttt{omega\_r} & $\Omega_R$ & radial frequency & $\mathrm{Gyr^{-1}}$ \\
\texttt{omega\_phi} & $\Omega_\phi$ & azimuthal frequency & $\mathrm{Gyr^{-1}}$ \\
\texttt{omega\_z} & $\Omega_z$ & vertical frequency & $\mathrm{Gyr^{-1}}$ \\
\texttt{theta\_r} & $\theta_R$ & radial angle & Rad. \\
\texttt{theta\_phi} & $\theta_\phi$ & azimuthal angle & Rad. \\
\texttt{theta\_z} & $\theta_z$ & vertical angle & Rad. \\
\texttt{rl} & $R_\mathrm{guide}$ & radius of a circular orbit at the same $L_z$ & kpc \\
\texttt{E} & $E$ & orbital energy & $\mathrm{km^2\ s^{-2}}$ \\
\texttt{EminusEc} & $E-E_{\mathrm{c}}$ & difference between orbit energy and energy of a circular orbit of the same $L_z$ &  $\mathrm{km^2\ s^{-2}}$ \\
\texttt{*\_err} & & error on each quantity & \\
\texttt{*\_*\_corr} & &correlation between the estimation of quantities & \\
\enddata
\tablecomments{Correlation coefficients are only computed for the orbital parameters and actions, and are computed separately for each set of quantities.}
\end{deluxetable*}

\subsection{A basic python example}\label{sec:use_python}

We now briefly demonstrate the use of the method in \texttt{galpy}. First we show the simplest implementation, which performs estimation within a \texttt{galpy.Orbit} instance in a given potential (in this case \texttt{MWPotential2014})

\begin{minted}{python}
from galpy.orbit import Orbit
from galpy.potential import MWPotential2014
#initialise an Orbit instance
o = Orbit(vxvv, pot=MWPotential2014)
eccentricity = o.e(analytic=True,type='staeckel')
rap = o.rap(analytic=True,type='staeckel')
rperi = o.rperi(analytic=True,type='staeckel')
zmax = o.zmax(analytic=True,type='staeckel')
\end{minted}

where \texttt{vxvv} gives the phase space coordinate of the object in question (\texttt{galpy} accepts various different forms of this input). Notice that it was not necessary to integrate the orbit before producing the estimates. It is also possible to run the estimation for large sets of phase space points using an instance of the \texttt{actionAngle.actionAngleStaeckel} class:

\begin{minted}{python}
from galpy.actionAngle import actionAngleStaeckel
aAS = actionAngleStaeckel(pot=MWPotential2014, 
			  delta=0.4)
e, zmax, rperi, rap = \
	aAS.EccZmaxRperiRap(R, vR, vT, z, vz, phi)
\end{minted}

where \texttt{R, vR, vT, z, vz, phi} are each an array of the corresponding coordinates, and each estimation is performed assuming $\Delta=0.4$. It is possible to calculate a separate $\Delta$ parameter for each of a set of objects by using the \texttt{estimateDeltaStaeckel} function

\begin{minted}{python}
from galpy.actionAngle import estimateDeltaStaeckel
delta = estimateDeltaStaeckel(mp, R, z, 
				no_median=True)
\end{minted}

these estimated $\Delta$ parameters can then be passed  to \texttt{aAS.EccZmaxRperiRap} using the \texttt{delta} keyword argument. The process can be sped up further by using the grid-based estimation method, which is implemented in \texttt{galpy} as \texttt{actionAngleStaeckelGrid}. A full tutorial which shows this implementation and the others listed here, as well as an example using real data, is available online\footnote{\url{http://galpy.readthedocs.io/en/latest/orbit.html}}.

\section{Conclusions}

We have demonstrated a new application of the \citet{2012MNRAS.426.1324B} St\"ackel fudge for the rapid calculation of the orbit parameters $r_\mathrm{peri}, r_\mathrm{ap}, z_\mathrm{max}$ and $e$, which does not depend on orbit integration. We have shown that for disc orbits, each parameter can generally be estimated to within less than a percent of the orbit integration value. We have demonstrated that this estimation is also valid outside the disc, but should be used cautiously for such orbits, where resonances can cause problems. We applied the method to the RAVE-TGAS data set of 216,201 stars, demonstrating the utility and speed of the method, which can return results as fast as $9\mathrm{\mu s}$ per object when used in its grid-based application. Thus, this technique can compute  point estimates for orbital parameters for, e.g., all $\approx 150$ million \emph{Gaia} stars with RVs at end of mission, in about 2.5 hr on a single CPU, which can be brought down to a few minutes of wall time because the computation can be trivially parallelised (It should be noted that this time is inflated by a factor of $\sim 100$ when uncertainties are propagated). We calculate the orbital parameters, as well as action-angle coordinates, frequencies and energies for the sample of stars with full 6D phase-space coordinates from \emph{Gaia} DR2, performing a full error propagation and estimation of the correlation between the parameters and the actions. This more robust estimation, which includes the estimation of the parameters for 100 realisations of the observed coordinates per star, takes $\sim 116$ hours of parallelised CPU time. 

Using the catalogue, which is available with the supplementary material, we show that the $r_\mathrm{peri}$-$r_\mathrm{ap}$ distribution for the \emph{Gaia} RV sample demonstrates the extensive `dynamical sampling' of the data, even when constrained to a relatively small sphere around the sun. We demonstrated that the clumpy features which are apparent in that space and the $r_\mathrm{mean}$-$e$ plane are coincident with regions of low mean $z_\mathrm{max}$, indicating that these features are likely the result of disk dynamical effects rather than halo dynamics. While angle-action variables may be a more fundamental label of stellar orbits, this tool may still be able to provide insight into the orbital structure of the Galaxy throughout the coming era of extremely large sets of stellar phase space coordinates. 

\section*{Acknowledgements}

The authors thank the anonymous reviewer for helpful reports, which improved the clarity and content of this paper. JTM acknowledges an STFC doctoral studentship, as well as the Dunlap Visitorship program and an RAS Grant, which funded an extended visit to the Dunlap Institute at the University of Toronto, to work on this project. JTM is also grateful for the hospitality of the Center for Computational Astrophysics at the Simons Foundation in New York City for a short period whilst working on this project. J.B. acknowledges the support of the Natural Sciences and Engineering Research Council of Canada (NSERC), funding reference number RGPIN-2015-05235, and from an Alfred P. Sloan Fellowship.

This work has made use of data from the European Space Agency (ESA) mission Gaia (~\url{http://www.cosmos.esa.int/gaia}), processed by the Gaia Data Processing and Analysis Consortium (DPAC, ~\url{http://www.cosmos.esa.int/web/gaia/dpac/consortium}). Funding for the DPAC has been provided by national institutions, in particular the institutions participating in the Gaia Multilateral Agreement. This project was developed in part at the 2018 NYC Gaia Sprint, hosted by the Center for Computational Astrophysics of the Flatiron Institute in New York City.

%% The reference list follows the main body and any appendices.
%% Use LaTeX's thebibliography environment to mark up your reference list.
%% Note \begin{thebibliography} is followed by an empty set of
%% curly braces.  If you forget this, LaTeX will generate the error
%% "Perhaps a missing \item?".
%%
%% thebibliography produces citations in the text using \bibitem-\cite
%% cross-referencing. Each reference is preceded by a
%% \bibitem command that defines in curly braces the KEY that corresponds
%% to the KEY in the \cite commands (see the first section above).
%% Make sure that you provide a unique KEY for every \bibitem or else the
%% paper will not LaTeX. The square brackets should contain
%% the citation text that LaTeX will insert in
%% place of the \cite commands.

%% We have used macros to produce journal name abbreviations.
%% \aastex provides a number of these for the more frequently-cited journals.
%% See the Author Guide for a list of them.

%% Note that the style of the \bibitem labels (in []) is slightly
%% different from previous examples.  The natbib system solves a host
%% of citation expression problems, but it is necessary to clearly
%% delimit the year from the author name used in the citation.
%% See the natbib documentation for more details and options.

\bibliography{bib}

%% This command is needed to show the entire author+affilation list when
%% the collaboration and author truncation commands are used.  It has to
%% go at the end of the manuscript.
%\allauthors

%% Include this line if you are using the \added, \replaced, \deleted
%% commands to see a summary list of all changes at the end of the article.
%\listofchanges

\end{document}